\pdfoutput=1
\pdfmapfile{+classico.map} 
\newif\ifafour
\afourtrue 
\newif\iftypodisclaim 
\typodisclaimtrue

\documentclass[\ifafour a4paper,12pt,\else a5paper,10pt,\fi
onecolumn,oneside,article,
british%
]{memoir}
\newif\ifpublic
\publictrue 

\newcommand*{\propertitle}{Unlearning and Seyab's theorem\\{\large a dialogue about updating probability}}
\newcommand*{\pdftitle}{Unlearning and Seyab's theorem: a dialogue about updating probability}
\newcommand*{\headtitle}{Unlearning and Seyab's theorem}
\newcommand*{\pdfauthor}{P.G.L.  Porta Mana}
\newcommand*{\headauthor}{\ifpublic Porta Mana%
\else\autanet\ Luca\fi}
\newcommand*{\reporthead}{}
\usepackage[T1]{fontenc} 
\input{glyphtounicode} \pdfgentounicode=1
\usepackage[utf8]{inputenx}

\usepackage{textcomp}
\usepackage[normalem]{ulem}
\usepackage{amsmath}
\usepackage{mathtools}
\usepackage{empheq}

\usepackage{framed}
\usepackage{amssymb}
\usepackage{amsxtra}

\usepackage[main=british,french,italian,german,swedish,latin,esperanto]{babel}

\usepackage[autostyle=false,autopunct=false,english=british]{csquotes}
\setquotestyle{american}

\usepackage{amsthm}

\theoremstyle{remark}

\newtheoremstyle{innote}{\parsep}{\parsep}{\footnotesize}{}{}{}{0pt}{}
\theoremstyle{innote}

\usepackage[shortlabels,inline]{enumitem}
\SetEnumitemKey{para}{itemindent=\parindent,leftmargin=0pt,listparindent=\parindent,parsep=0pt,itemsep=\topsep}
\setlist[enumerate,2]{label=\alph*.}
\setlist[enumerate]{leftmargin=\parindent}
\setlist[itemize]{leftmargin=\parindent}
\setlist[description]{leftmargin=\parindent}

\usepackage[babel,theoremfont]{newpxtext}
\usepackage[bigdelims,nosymbolsc
]{newpxmath}
\useosf
\makeatletter
\def\re@DeclareMathSymbol#1#2#3#4{%
    \let#1=\undefined
    \DeclareMathSymbol{#1}{#2}{#3}{#4}}
\re@DeclareMathSymbol{\bigoplusop}{\mathop}{largesymbols}{"4C}
\re@DeclareMathSymbol{\bigotimesop}{\mathop}{largesymbols}{"4E}
\re@DeclareMathSymbol{\sumop}{\mathop}{largesymbols}{"50}
\re@DeclareMathSymbol{\prodop}{\mathop}{largesymbols}{"51}
\re@DeclareMathSymbol{\bigcupop}{\mathop}{largesymbols}{"53}
\re@DeclareMathSymbol{\bigcapop}{\mathop}{largesymbols}{"54}
\re@DeclareMathSymbol{\bigwedgeop}{\mathop}{largesymbols}{"56}
\re@DeclareMathSymbol{\bigveeop}{\mathop}{largesymbols}{"57}
\re@DeclareMathSymbol{\bigtimesop}{\mathop}{largesymbolsPXA}{"10}
\makeatother

\DeclareFontFamily{U}{egreek}{\skewchar\font'177}%
\DeclareFontShape{U}{egreek}{m}{n}{<-6>s*[1]eurm5 <6-8>s*[1]eurm7 <8->s*[1]eurm10}{}%
\DeclareFontShape{U}{egreek}{m}{it}{<->s*[1]eurmo10}{}%
\DeclareFontShape{U}{egreek}{b}{n}{<-6>s*[1]eurb5 <6-8>s*[1]eurb7 <8->s*[1]eurb10}{}%
\DeclareFontShape{U}{egreek}{b}{it}{<->s*[1]eurbo10}{}%
\DeclareSymbolFont{egreeki}{U}{egreek}{m}{it}%
\SetSymbolFont{egreeki}{bold}{U}{egreek}{b}{it}
\DeclareSymbolFont{egreekr}{U}{egreek}{m}{n}%
\SetSymbolFont{egreekr}{bold}{U}{egreek}{b}{n}
\DeclareFontFamily{U}{egreekx}{\skewchar\font'177}
\DeclareFontShape{U}{egreekx}{m}{n}{%
       <-7.5>s*[0.9]euex7%
    <7.5-8.5>s*[0.9]euex8%
    <8.5-9.5>s*[0.9]euex9%
    <9.5->s*[0.9]euex10%
}{}
\DeclareSymbolFont{egreekx}{U}{egreekx}{m}{n}
\DeclareMathSymbol{\sumop}{\mathop}{egreekx}{"50}
\DeclareMathSymbol{\prodop}{\mathop}{egreekx}{"51}
\DeclareMathSymbol{\coprodop}{\mathop}{egreekx}{"60}
\makeatletter
\def\sum{\DOTSI\sumop\slimits@}
\def\prod{\DOTSI\prodop\slimits@}
\def\coprod{\DOTSI\coprodop\slimits@}
\makeatother
 \DeclareMathSymbol{\partialup}{\mathalpha}{egreekr}{"40}
 \DeclareMathSymbol{\epsilon}{\mathalpha}{egreeki}{"0F}
 \DeclareMathSymbol{\kappa}{\mathalpha}{egreeki}{"14}
 \DeclareMathSymbol{\lambda}{\mathalpha}{egreeki}{"15}
 \DeclareMathSymbol{\nu}{\mathalpha}{egreeki}{"17}
 \let\varkappa\kappa
 \DeclareMathSymbol{\varAlpha}{\mathalpha}{egreeki}{"41}
 \DeclareMathSymbol{\varBeta}{\mathalpha}{egreeki}{"42}
 \DeclareMathSymbol{\varDelta}{\mathalpha}{egreeki}{"01}
 \DeclareMathSymbol{\varEpsilon}{\mathalpha}{egreeki}{"45}
 \DeclareMathSymbol{\varIota}{\mathalpha}{egreeki}{"49}
 \DeclareMathSymbol{\varNu}{\mathalpha}{egreeki}{"4E}
 \DeclareMathSymbol{\varOmicron}{\mathalpha}{egreeki}{"4F}
 \DeclareMathSymbol{\deltaup}{\mathalpha}{egreekr}{"0E}
  \DeclareMathSymbol{\piup}{\mathalpha}{egreekr}{"19}

\renewcommand\sfdefault{uop}
\DeclareMathAlphabet{\mathsf}  {T1}{\sfdefault}{m}{sl}
\SetMathAlphabet{\mathsf}{bold}{T1}{\sfdefault}{b}{sl}

\usepackage[scaled=0.84]{DejaVuSansMono}

\usepackage{mathdots}

\usepackage[usenames]{xcolor}
\definecolor{mybluishpurple}{RGB}{51,34,136}
\definecolor{myblue}{RGB}{136,204,238}
\definecolor{mybluishgreen}{RGB}{68,170,153}
\definecolor{mygreen}{RGB}{17,119,51}
\definecolor{mygreenishyellow}{RGB}{153,153,51}
\definecolor{myyellow}{RGB}{221,204,119}
\definecolor{myred}{RGB}{204,102,119}
\definecolor{mypurplishred}{RGB}{136,34,85}
\definecolor{myreddishpurple}{RGB}{170,68,153}
\definecolor{mygrey}{RGB}{221,221,221}
\colorlet{shadecolor}{mygrey}

\usepackage{bm}
\usepackage{microtype}

\usepackage[backend=biber,mcite,
citestyle=authoryear-comp,bibstyle=pglpm-authoryear,autopunct=false,sorting=ny,sortcites=false,natbib=false,maxcitenames=1,maxbibnames=8,minbibnames=8,giveninits=true,uniquename=false,uniquelist=false,maxalphanames=1,block=space,hyperref=true,defernumbers=false,useprefix=true,sortupper=false,language=british,parentracker=false]{biblatex}
\DeclareSortingScheme{ny}{\sort{\field{sortname}\field{author}\field{editor}}\sort{\field{year}}}
\DefineBibliographyExtras{british}{\def\finalandcomma{\addcomma}}
\renewcommand*{\finalnamedelim}{\addcomma\space}
\DeclareDelimFormat{multicitedelim}{\addsemicolon\space}
\DeclareDelimFormat{compcitedelim}{\addsemicolon\space}
\DeclareDelimFormat{postnotedelim}{\space}
\renewcommand{\bibfont}{\footnotesize}
\defbibheading{bibliography}[\bibname]{\section*{#1}\addcontentsline{toc}{section}{#1}
}
\newcommand*{\citep}{\parencites}
\newcommand*{\citey}{\parencites*}
\renewcommand*{\cites}{\parencites}
\providecommand{\href}[2]{#2}

\newcommand*{\amp}{\&}

\newcommand*{\subtitleproc}[1]{}

\usepackage{graphicx}
\usepackage{wrapfig}

\usepackage[hypertexnames=false]{hyperref}
\usepackage[depth=4]{bookmark}
\hypersetup{colorlinks=true,bookmarksnumbered,pdfborder={0 0 0.25},citebordercolor={0.2 0.1333 0.5333},
citecolor=mybluishpurple,linkbordercolor={0.0667 0.4667 0.2},
linkcolor=mypurplishred,urlbordercolor={0.5333 0.1333 0.3333},
urlcolor=mygreen,breaklinks=true,pdftitle={\pdftitle},pdfauthor={\pdfauthor}}

\ifafour\setstocksize{297mm}{210mm}
\else\setstocksize{210mm}{5.5in}
\fi%
\settrimmedsize{\stockheight}{\stockwidth}{*}%
\setlxvchars[\normalfont] 
\setxlvchars[\normalfont]%
\ifafour\settypeblocksize{*}{32pc}{1.618}
\else\settypeblocksize{*}{26pc}{1.618}
\fi%
\setulmargins{*}{*}{1}
\setlrmargins{*}{*}{*}%
\setheadfoot{\onelineskip}{2.5\onelineskip}%
\setheaderspaces{*}{2\onelineskip}{*}%
\setmarginnotes{2ex}{10mm}{0pt}%
\checkandfixthelayout[nearest]%
\fixpdflayout%
\pdfmapline{+dummy-space <dummy-space.pfb}\pdfinterwordspaceon%

\newenvironment{acknowledgements}{\section*{Thanks}\addcontentsline{toc}{section}{Thanks}}{\par}
\makeatletter\renewcommand{\appendix}{\par
  \bigskip{\centering
   \interlinepenalty \@M
   \normalfont
   \printchaptertitle{\sffamily\appendixpagename}\par}
  \setcounter{section}{0}%
  \gdef\@chapapp{\appendixname}%
  \gdef\thesection{\@Alph\c@section}%
  \anappendixtrue}\makeatother
\counterwithout{section}{chapter}
\setsecnumformat{\upshape\csname the#1\endcsname\quad}
\setsecheadstyle{\large\bfseries\sffamily%
\raggedright}
\setsubsecheadstyle{\bfseries\sffamily%
\raggedright}
\setbeforesecskip{-2ex plus 1ex minus .2ex}
\setaftersecskip{0.5ex plus .2ex }
\setaftersubsecskip{-1em}
\setsubsecindent{0pt}
\setparaheadstyle{\bfseries\sffamily%
\raggedright}
\newcommand{\addchap}[1]{\chapter*[#1]{#1}\addcontentsline{toc}{chapter}{#1}}
\newcommand{\addsec}[1]{\section*{#1}\addcontentsline{toc}{section}{#1}}

\copypagestyle{manaart}{plain}
\makeheadrule{manaart}{\headwidth}{0.5\normalrulethickness}
\makeoddhead{manaart}{%
{\footnotesize
\scshape\headauthor}}{}{{\footnotesize\sffamily%
\headtitle}}
\makeoddfoot{manaart}{}{\thepage}{}
\newcommand*\autanet{\includegraphics[height=\heightof{M}]{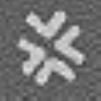}}
\definecolor{mygray}{gray}{0.333}
\iftypodisclaim%
\ifafour
\makeoddfoot{plain}{}{\makebox[0pt]{\thepage}}{}%
\else\newcommand\addprintnote{\begin{picture}(0,0)%
\put(176,112){\makebox(0,0){\rotatebox{90}{\tiny\color{mygray}\textsf{This
            document is designed for screen reading and
            two-up printing on A4 or Letter paper}}}}%
\end{picture}}%
\makeoddfoot{plain}{}{\makebox[0pt]{\thepage}\addprintnote}{}%
\fi
\else
\makeoddfoot{plain}{}{\makebox[0pt]{\thepage}}{}
\fi
\makeoddhead{plain}{}{}{\footnotesize\reporthead}

\pretitle{\begin{center}\LARGE\sffamily%
\bfseries}
\posttitle{\bigskip\end{center}}

\makeatletter\newcommand*{\atf}{\includegraphics[
totalheight=\heightof{@}]{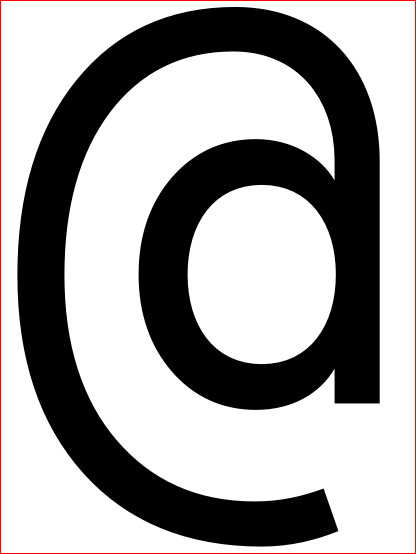}}\makeatother

\providecommand{\epost}[1]{\texttt{\footnotesize\textless#1\textgreater}}
\providecommand{\email}[2]{\href{mailto:#1ZZ@#2 ((remove ZZ))}{#1\protect\atf#2}}

\preauthor{\vspace{-0.5\baselineskip}\begin{center}
\normalsize\sffamily%
\lineskip  0.5em}
\postauthor{\par\end{center}}
\predate{\DTMsetdatestyle{mydate}\begin{center}\footnotesize}
\postdate{\end{center}\vspace{-\medskipamount}}
\usepackage[british]{datetime2}
\DTMnewdatestyle{mydate}%
{
}
\DTMsetdatestyle{mydate}

\setfloatadjustment{figure}{\footnotesize}
\captiondelim{\quad}
\captionnamefont{\footnotesize\sffamily%
}
\captiontitlefont{\footnotesize}
\firmlists*
\midsloppy

\title{\propertitle
}
\author{\ifpublic%
P.G.L. Porta\,Mana%
\else Luca\fi
\quad
\epost{\email{mana}{kth.se}}%
}

\date{\today}

\newcommand*{\p}{\mathrm{P}}
\renewcommand*{\|}{\mathpunct{|}}
\newcommand*{\sect}{\S}
\newcommand*{\eg}{{e.g.}}

\newcommand*{\speak}{\smallskip\textbf}
\newcommand*{\sa}{\speak{Salv.} }
\newcommand*{\si}{\speak{Simp.} }
\newcommand*{\yA}{\varAlpha}
\newcommand*{\yB}{\varBeta}
\newcommand*{\yI}{\varIota}
\newcommand*{\yO}{\varOmicron}
\newcommand*{\yT}{\varOmicron'}
\newcommand*{\yN}{\varNu}
\newcommand*{\epsi}{\epsilon}

\firmlists
\begin{document}
\captiondelim{\quad}\captionnamefont{\footnotesize}\captiontitlefont{\footnotesize}
\selectlanguage{british}\frenchspacing

\maketitle

\selectlanguage{british}\frenchspacing
 \vspace{-\bigskipamount}

\setlength{\epigraphwidth}{.65\columnwidth}
\epigraphtextposition{flushright}
\epigraphfontsize{\footnotesize}
\setlength{\epigraphrule}{0pt}
\epigraph{\enquote{You can never know everything,} Lan said quietly, \enquote{and part of what you know is always wrong. Perhaps even the most important part. A portion of wisdom lies in knowing that. A portion of courage lies in going on anyway.}}{\citep{jordan2000_r2010}}


\smallskip
\noindent \emph{The interlocutors Salviati and Simplicio are two students very passionate about the
  probability calculus, logic, and philosophy.}
\smallskip

\sa It's  beautifully simple how Bayes's theorem,
\begin{equation}
  \label{eq:bayes}
  \p(\yB \| \yA \land \yI) = \frac{\p(\yA \| \yB \land \yI) \; \p(\yB \| \yI)}{\p(\yA \| \yI)}
\end{equation}
relates the probabilities we give to a proposition $\yB$ in the two cases:
when we know only $\yI$, and when we also know $\yA$ besides $\yI$. Isn't it,
Simplicio?

\si It surely is, Salviati. Don't forget that the theorem requires you to
specify either $\p(\yA \| \yB \land \yI)$ and $\p( \yA \land \yI)$, or
$\p(\yA \land \yB \| \yI)$ and $\p( \yA \| \lnot \yB \land \yI)$, or some
similar combination, to get the latter of the two probabilities you mention
from the former.

\sa Of course. And as you just said, we usually calculate
$\p(\yB \| \yA \land \yI)$ given $\p(\yB \| \yI)$, don't we?

\si Yes, that's quite natural, because the theorem tells us how to update
our probability about $\yB$ when we learn $\yA$. That's why we call
$\p(\yB \| \yA \land \yI)$ the \emph{posterior} probability and
$\p(\yB \| \yI)$ the \emph{prior} probability.

\sa You see, I was thinking about something curious connected with this
\emph{prior} and \emph{posterior} business. In our lives we often learn new
things and acquire new knowledge. But at times we also unlearn, forget,
lose data, or discover that some of our knowledge is unfounded. Don't we?

\si Absolutely! Something like that happened to me recently. Do you
remember what they told us in our classes in quantum mechanics -- that if
you prepare a particle with a very precise position, then a measurement
of its momentum will give a completely uncertain result?

\sa I remember. Of course, position and momentum are conjugate quantities.

\si Well, what our teachers said isn't really true. It turns out that if
you make \emph{one} measurement of the momentum of that particle, you can 
get a definite result, as precise as you like.

\sa Heresy!

\si It's true and actually very sensible. Consider a simpler case: you
prepare an electron with spin in the positive $x$ direction, say, and then
make \emph{one} measurement of its spin in the $y$ direction. The result of
the measurement is one definite spot on the screen at the exit of the
measuring Stern-Gerlach apparatus \citep[fig.~1.5, \sect~1-5,
p.~14]{peres1995}: we know precisely the result, either $+\hbar/2$ or
$-\hbar/2$ in the $y$ direction. Yet the measurement error, according to
the standard formulae, is $\hbar/2$.

\sa That sounds true. But then how do you reconcile this with the
formulae? Were our teachers lying?

\si The point is that the formulae refer to the statistics of \emph{many}
measurements, not just one. If you repeat the spin measurement with the
same experimental setup, you'll observe a new and precise result, either
$+\hbar/2$ or $-\hbar/2$. As you repeat this kind of experiment, both
results will be observed in about a fifty-fifty ratio. The
\enquote{measurement error} refers to the standard deviation in \emph{many}
measurements. The situation with position and momentum is analogous. Our
teachers were not lying but using a too vague or slightly wrong terminology
about what the \enquote{measurement error} was. Read Ballentine
\citey{ballentine1970}, or take a look at the experimental results in
Leonhardt \citey{leonhardt1997} -- \eg\ his fig.~2.1, \sect~2.2, p.~23 --
to have a better understanding of the situation. No paradoxes.

\sa Amazing, it looks like I have also unlearnt something today.

\si As you said, it happens. But how does this relate to Bayes's theorem?

\sa Right. You see, I was thinking about the three rules of the probability calculus,
\begin{equation}
  \label{eq:rules_prob}
  \begin{gathered}
  \p(\lnot \yA \|\yI) = 1 - \p(\yA \| \yI),\qquad
  \p(\yB \land \yA \|\yI) = \p(\yB \| \yA \land \yI)\; \p(\yA \| \yI),\\
  \p(\yA \lor \yB \|\yI) = \p(\yB \| \yI) +  \p(\yA \| \yI) - \p(\yB \land \yA \|\yI).
\end{gathered}
\end{equation}
These rules, combined, give us relations among probabilities with different
conditionals. By the way, do you mind if I use the term \emph{context} to
refer to the conditional?

\si It isn't standard terminology I think, but it sounds like a sensible
term. Maybe you could even use \emph{situation}, similarly to Barwise's
\citey{barwise1989} notion in Logic; I've been wondering for a long time
whether Barwise's notion of \emph{situation} in Logic has deeper
connections with probability conditionals. But that's a discussion topic
for another time. Please go on.

\sa Thank you. Bayes's theorem is an example of what I was saying about
relations between contexts: it relates the probabilities for $\yB$ in the
context where we know $\yI$ and in the context where we know
$\yA \land \yI$. Now here's my point: the probability rules tell us nothing
about which of these two contexts comes first in time; they don't say which
should be \emph{prior} and which \emph{posterior}. Now imagine a situation
in which we have knowledge of $\yA \land \yI$, but some time later we
forget whether $\yA$ is true. Then $\p(\yB \| \yA \land \yI)$ would be our
\emph{prior} probability, and $\p(\yB \| \yI)$ our \emph{posterior}.

\si Mmm\ldots\ In that case wouldn't we also forget the numerical value of
$\p(\yB \| \yA\land \yI)$? Or, if we have saved the value of this
probability somewhere, it would remind us that we knew $\yA$ was true, so
we should still need to consider $\p(\yB \| \yA \land \yI)$, shouldn't we?

\sa Right, my example wasn't very logical. Consider this then: we are
initially sure about $\yA$, besides $\yI$, so we use probabilities with the
context $\yA \land \yI$. But then we somehow discover that $\yA$ isn't
actually certain; our certainty was unfounded. Upon this discovery we
should start using probabilities with context $\yI$ only. Our probability
for $\yB$ should be updated in reverse, using \enquote{Seyab's theorem}:
\begin{equation}
  \label{eq:seyab}
  \p(\yB \| \yI) = \p(\yA \| \yI)\,\frac{  \p(\yB \| \yA \land \yI) }{\p(\yA \| \yB \land \yI)},
\end{equation}
which expresses our \emph{unlearning} of $\yA$. The probability
$\p(\yB \| \yA \land \yI)$ is our \emph{prior}, and $\p(\yB \| \yI)$ our
\emph{posterior}. The probability $\p(\yA \| \yI)$ appears as a
normalization factor just like in Bayes's theorem.

\si I feel that there's an inconsistency in your example. If the context
$\yI$ doesn't give an extremal, $0$ or $1$ probability to $\yA$, then that
context does not contradict the possibility that $\yA$ might be true. So in
that context there's the possibility that we discover later that $\yA$ is
true, and therefore move $\yA$ into the context. But once $\yA$ enters the
context, and we use the new context $\yA \land \yI$, the possibility that
$\yA$ might be uncertain contradicts this new context. Hence we can't
revert to using $\yI$ alone.

\sa Yours is an interesting assertion. You seem to be invoking an
additional rule on how to use the three probability rules. Do the three
rules~\eqref{eq:rules_prob} say that you can't take a proposition out of a
context, once you're using that context for your plausibility judgements?
Where or how do the rules exactly say this?

\si Well, the probability rules don't prescribe which context we should be
using, for that matter. I think it's implicit in their use that the context
of probability judgement should reflect our state of knowledge or whatever
working hypotheses we're entertaining. Just like in truth logic we use
axioms that express what we know, when we want to reason out a conclusion
deductively.

\sa I agree with the point of view you just expressed. But I think it
endorses what I was saying: if we are initially sure about $\yA$ and $\yI$,
then we should use the context $\yA\land \yI$; if we later become uncertain
about $\yA$, then we should switch to the context $\yI$. I thought you
agreed with me that in real life we can happen to become unsure about
something we were previously sure of.

\si Your reasoning appears formally correct, but I still can't shake off
the feeling that it contains a contradiction. OK, Let me change my
argument a little. I think that we shouldn't have used the context
$\yA \land \yI$ in the first place. You say: \enquote{we're initially
  certain about $\yA$}; but I rather imagine that we initially didn't know
$\yA$ directly: we had some other kind of knowledge instead, say expressed
by the proposition $\yO$, that gave almost unit probability to $\yA$:
\begin{equation}
  \label{eq:almost_certain_A}
  \p(\yA \| \yO \land \yI) = 1- \epsi, \quad \epsi \ll 1,
\end{equation}
and, by the probability rules,
\begin{equation}
  \label{eq:prob_B_A_indirectly_certain}
  \begin{aligned}
    \p(\yB \| \yO \land \yI) &=
                               \!\begin{multlined}[t][0.6\linewidth]
                                 \p(\yB \| \yA \land \yO \land \yI)\;\p(\yA \| \yO \land \yI) +{}\\
                                 \p(\yB \| \lnot\yA \land \yO \land \yI)\;\p(\lnot\yA \| \yO \land \yI),
\end{multlined}
\\
  &\approx \p(\yB \| \yA \land \yO \land \yI).
  \end{aligned}
\end{equation}
So $\yO \land \yI$,  not $\yA \land \yI$, was our initial context,
leading to probability judgements very similar to assuming $\yA$ to be
true. When you say \enquote{we discover that $\yA$ is uncertain}, I imagine
that we learn some new fact, say expressed by $\yT$, that makes $\yA$ again
very uncertain, even nullifying our knowledge of $\yO$:
\begin{equation}
  \label{eq:again_uncertain_A}
  \p(\yA \| \yT \land \yO \land \yI) = \p(\yA \|  \yI),
\end{equation}
and in this new context, $\yT \land \yO \land \yI$, our probability
judgement for $\yB$ is numerically very close to $\p(\yB \| \yI)$. I think
this analysis gives numerical values similar to yours, but is closer to
what would happen in a real situation of \enquote{unlearning}.

\sa Your analysis is absolutely valid and can be an accurate representation
of many a real situation. But I think that it lends a hand to my own
argument. First, I can still imagine a situation in which we become unsure
about $\yO$, the proposition that you assumed to be certain. In this case I
suppose we could make an analysis similar to the one you just made, but
with $\yO$ in place of $\yA$, introducing some new proposition $\yN$
relating to $\yO$ as $\yO$ related to $\yA$. But this game could go on
forever.

\si You seem to be saying that indubitable, certain knowledge is impossible
to acquire. I completely disagree with such philosophical position.

\sa That wasn't my intention. My point is to show that your analysis
doesn't deny the possibility that we might become unsure about something we
were sure of. Your analysis doesn't show any contradiction in my position;
it only sidesteps it.

\si That's true. I feel the need to sidestep it because, to me, saying that
$\yA$ is certain seems to exclude the possibility that we can become
uncertain about it. Doesn't this possibility contradict the notion of
\enquote{certainty} itself? wouldn't it mean that we were \emph{not}
completely certain about $\yA$ in the first place?

\sa That's an important philosophical and semantic question. I partly
understand what you mean. And yet I'm not convinced that being sure about
something at some time contradicts being unsure about it some time later.
In any case, what's important is that the probability rules make allowance
for this possibility; don't they? Seyab's theorem, derived from them,
seems to suggest exactly this possibility.

\si I agree, I don't have a philosophical proof of my view about
\enquote{certainty}. To be honest I'm not acquainted with any philosophical
works about this concept. I was just giving voice to my common sense; but
it seems yours is at variance with mine. My analysis shows that we can make
sense of your \enquote{unlearning}, though, by adding more propositions to the
context rather than by deleting them.

\sa The numerical probability values of your analysis are practically
identical with the ones we would obtain with Seyab's
theorem~\eqref{eq:seyab}. Or at least, your analysis doesn't show that
direct application of Seyab's theorem would lead to inconsistent values. So
why can't we just apply it? Why can't we simply move $\yA$ out of the
context?

\si With some more thinking I might find a contradiction in your view. But
for the moment let me ask you a different question: in which situations
could your Seyab theorem be used? Do you have any concrete applications in
mind?

\sa No, I don't. As I said, it was a curiosity and prompted my question
about updating contexts. But we live in an age when knowledge and
information are stored at a rate never experienced before. For this reason,
loss and deterioration of knowledge and information become more likely,
too. The outcome of a laboratory experiment could be misrecorded, the
results of a medical test could be lost, a meteorological report could be
misinterpreted. Couldn't Seyab's theorem turn out to be useful in such
cases?

\si I don't think it likely, but as a scientist I'll give you the benefit
of doubt of course. We never know.

\setlength{\intextsep}{0.5ex}

\ifpublic
\begin{acknowledgements}
  \ldots to Jakob Jordan and Alper Yegenoglu for many insightful
  discussions about probability conditionals. To Mari \amp\ Miri for
  continuous encouragement and affection. To Buster Keaton and Saitama
  for filling life with awe and inspiration. To the developers and
  maintainers of \LaTeX, Emacs, AUC\TeX, Open Science Framework, biorXiv,
  Hal archives, Python, Inkscape, Sci-Hub for making a free and
  unfiltered scientific exchange possible.
\sourceatright{\autanet}
\end{acknowledgements}
\fi


\defbibnote{prenote}{{\footnotesize (\enquote{de $X$} is listed under D,
    \enquote{van $X$} under V, and so on, regardless of national
    conventions.)\par}}

\printbibliography[
]

\end{document}